\documentclass[superscriptaddress,twocolumn,prl,showpacs,floatfix]{revtex4}
\usepackage[dvips]{graphicx}
\usepackage{epsfig}
\usepackage{psfrag}
\usepackage{amsmath,amsfonts,amssymb,bm,ulem}
\usepackage[usenames]{color}

\begin{document}
\title{Effect of triple interaction on energy delocalization in the strongly disordered system of interacting two-level defects. Comment to the recent work ``Interaction driven relaxation of two level systems in glasses'' by D. Bodea and A. Wurger}
\author{A. L. Burin} 
\affiliation{Pacific Institute of Theoretical Physics,
University of British Columbia,
6224 Agricultural Road,
Vancouver, BC
Canada, V6T 1Z1}
\affiliation{Department of Chemistry, Tulane University, New
Orleans, LA 70118, USA}
\author{I. Ya. Polishchuk}
\affiliation{Department of Chemistry, Tulane University, New
Orleans, LA 70118, USA}
\begin{abstract}
We accurately treat the effect of the double and triple interactions of two-level systems (TLS) in glasses onto the energy delocalization due to the long-range  interaction of TLS. Although this work qualitatively reproduces the estimates of our previous work \cite{BMP} we believe that it is important because in this paper the estimates are done with quantitative accuracy. This work can serve as the important step towards the development of quantitative theory of many-body delocalization due to the long-range interaction. Our study is compared with the recent work by Bodea et al \cite{subject} claiming that the interaction of TLS triples leads to the energy delocalization. We cannot agree with the mentioned work because as we show the interaction of TLS triples was overestimated there and therefore the result for the TLS relaxation rate obtained there is invalid. This work serves as the extended complementary version of the one page comment submitted to Physical Review Letters.  
\end{abstract}

\pacs{73.23.-b 72.70.+m 71.55.Jv 73.61.Jc 73.50.-h 73.50.Td}

\maketitle

The problem of Anderson localization in the presence of an interaction attracts a large attention since the concept of localization has been suggested in the classical work.\cite{pwa} Particularly, the long-range interaction $1/R^{a}$ with a sufficiently small exponent $a$ ($a\leq 3$)  can lead to the delocalization at arbitrarily strong disordering.\cite{pwa,Levitov} The ensemble of two level systems (TLS) in amorphous solids \cite{ahwp} is  the particularly interesting system to study the effect of the long-range  many-body interaction $J(R) \sim U_{0}/R^3$ on the localization, because the interaction is extremely weak their, i. e. the dimensionless product of the density of two level system states $P_{0}$ and their interaction constant $U_{0}$, is the universally small value around $10^{-3}$.\cite{Hunklinger} After the experimental demonstration that the low temperature TLS relaxation rate shows linear temperature dependence \cite{pablo,echo,osheroff} instead of $T^{3}$ dependence due to the TLS-phonon interaction we propose the scenario,  where this anomaly was explained by the TLS interaction stimulated delocalization leading to the irreversible  relaxation.\cite{BMP}

Recently the model of a TLS interaction stimulated relaxation has been considered by Bodea et al.\cite{subject} They obtained the estimate of the TLS relaxation rate different from Ref. \cite{BMP} both qualitatively and quantitatively. The method used in Ref. \cite{subject} was based on the analysis  of resonant couplings developed in our previous work.\cite{BMP} The number of resonant couplings of single TLS with surrounding TLS (resonant pairs) and TLS pairs (resonant triples) were calculated. The resonant coupling was defined as in Refs. \cite{BMP,Levitov}, namely two entities $A$ and $B$ are in resonance, when their joint transition amplitude  $V_{AB}$ induced by their interaction exceeds the difference between their energies $E_{A}-E_{B}$ calculated ignoring their interaction. The criterion for energy delocalization was chosen using the condition of Ref. \cite{BMP} that the given TLS have more than one resonant interactions. 

The number of resonant pairs $W_{2}$ per the single TLS was found to be very small \cite{subject} in full agreement with Ref. \cite{BMP} where this estimate was already made. Below we reproduce this estimate to save time of the reader. The Hamiltonian of two interacting TLS can be expressed as 
\begin{eqnarray}
-\Delta_{i}S_{i}^{z}-\Delta_{j}S_{j}^{z}
-\Delta_{0i}S_{i}^{x}-\Delta_{0j}S_{j}^{x}
-U_{ij}S_{i}^{z}S_{j}^{z}. 
\label{eq:Hdouble}
\end{eqnarray} 
Then we can develop analytical approach for double transitions treating TLS interaction as a perturbation. To develop the perturbation theory with respect to the interaction we diagonalize the noninteracting Hamiltonian introducing new spin operators 
\begin{eqnarray}
s_{i}^{z}=\frac{\Delta_{i}}{E_{i}}S_{i}^{z}+\frac{\Delta_{0i}}{E_{i}}S_{i}^{x},~ s_{i}^{x}=-\frac{\Delta_{0i}}{E_{i}}S_{i}^{z}+\frac{\Delta_{i}}{E_{i}}S_{i}^{x},
\nonumber\\
E_{i,j}=\sqrt{\Delta_{i,j}^{2}+\Delta_{0i,j}^{2}}
\label{eq:new_spins}
\end{eqnarray}
Then the Hamiltonian of pair can be rewritten as
\begin{widetext}
\begin{eqnarray}
\widehat{H}_{pair}=\widehat{H}_{0}+\widehat{U},
\widehat{H}_{0}=-E_{i}s_{i}^{z}-E_{j}s_{j}^{z}, ~ \widehat{U}=-\frac{1}{2}U_{ij}\frac{\Delta_{i}s_{i}^{z}-\Delta_{0i}s_{i}^{x}}{E_{i}}\frac{\Delta_{j}s_{j}^{z}-\Delta_{0j}s_{j}^{x}}{E_{j}}.
\label{eq:Hdouble1}
\end{eqnarray} 
\end{widetext}
The basis states of the Hamiltonian $H_{0}$ can be represented in terms of the projections of spins $s_{i}^{z}=\pm s_{j}^{z} =\pm 1/2 =\mid \pm>$. The transition amplitude of the pair $|+-> \rightarrow |-+>$ is defined by the as the corresponding matrix element of interacting Hamiltonian 
\begin{eqnarray}
J_{ij}=\frac{1}{4}U_{ij}\frac{\Delta_{0i}\Delta_{0j}}{E_{i}E_{j}}. 
\label{eq:Hdoubletr1}
\end{eqnarray} 
The resonant pair is formed when 
\begin{eqnarray}
\mid E_{i}-E_{j}\mid < J_{ij}. 
\label{eq:Hdoubleres}
\end{eqnarray} 
The probability of resonance for the single pair can be estimated as $g\mid J_{ij}\mid$, where $g$ is the width of the distribution of a given TLS over energies. 
Then we need to sum all probabilities over all neighbouring TLS $j$. This sum can be expressed as the integral 
\begin{eqnarray}
W_{2}\approx P_{0}\int\frac{d\Delta_{0j}}{\Delta_{0j}} \int d{\bf r}_{j}\mid J_{ij}\mid
\nonumber\\
\theta(\lambda_{T}-r_{j})\theta(r_{j}-a_{T})
\nonumber\\
\approx P_{0}U_{0}\ln\left(\frac{k_{B}T}{\hbar \gamma_{ph}}\right)\ll 1,  
\label{eq:double_res_est}
\end{eqnarray} 
where the distance $r_{j}$ is taken in the range between their minimum value $a_{T}\sim (U_{0}/T)^{1/3}$ and the resonant phonon wavelength $\lambda_{T}=c/T$ where $c$ is the sound velocity. The distance cannot exceed the minimum distance because at shorter distances the interaction between TLS exceeds their thermal energy and therefore it is a very small probability to find any of them in the excited state. The upper constraint is because the interaction length is always restricted to the resonant wavelength. We also use the notation $\gamma_{ph} \sim U_{0}/(\lambda_{T}^{3}\hbar)$ corresponding to the TLS relaxation time associated with their interaction constant as in \cite{subject}.

However, in contrast with \cite{BMP} Bodea et al found that each TLS has the large number of triple resonant interactions with surrounding pairs of other TLS. According to \cite{subject} this should lead to the delocalization of TLS energy and the irreversible relaxation of each TLS. 
On our opinion the estimate of the number of resonant triples $W_{3}$ made in Ref. \cite{subject} exceeds the real value because the authors overestimated  the triple TLS transition amplitude $J$ assuming 
\begin{equation}
J^{2}=\left(\Delta_{i}/E_{i}\right)^{2}J_{ik}^{2}+\left(\Delta_{i}/E_{i}\right)^{2}J_{jk}^{2}. 
\label{eq:bodea} 
\end{equation}
Here  $\Delta$ and $E=\sqrt{\Delta^{2}+\Delta_{0}^{2}}$ are asymmetry energy and energy of TLS,  while $\Delta_{0}$ is its tunnelling amplitude. The expression Eq. (\ref{eq:bodea}) for the triple transition amplitude $J$ leads to the result, which conflicts with the common sense, that if the TLS $j$ is located arbitrarily far from TLS $i$ and $k$ ($R_{ij}\approx R_{jk} \gg R_{ik}$) then the amplitude $J$ does not depend on the long distance $R_{ij}$ and is defined only by the short distance $R_{ik}$ leading to the result $J \approx \overline{u}_{i}J_{ik} \sim U_{0}/R_{ik}^{3}$. On our opinion the triple TLS transition amplitude should vanish with vanishing the coupling between far separated entities. To prove that we perform  analytical and numerical calculations of the triple transition amplitude and estimate the number of resonant triples that turns out to be much less then one. 

To calculate the triple transition amplitude we introduce the triple TLS pseudospin Hamiltonian 
\begin{equation}
\widehat{H}_{triple}=-\sum_{i=1}^{3}\Delta_{i}S_{i}^{z}-\sum_{i=1}^{3}\Delta_{0i}S_{i}^{x}-\frac{1}{2}\sum_{i,j=1}^{3}U_{ij}S_{i}^{z}S_{j}^{z}. 
\label{eq:Htriple}
\end{equation} 
The long-range interaction $U_{ij} \approx U_{0}/R_{ij}^{3}$ essentially contributes to the delocalization for the far separated TLSs so we assume that the interaction is smaller than all TLS energies $E = \sqrt{\Delta^{2}+\Delta_{0}^2}$. This condition is satisfied in all previous studies for relevant TLS clusters.\cite{BMP,subject} Then we can develop analytical approach for triple transitions treating TLS interaction as a perturbation. To develop the perturbation theory with respect to the interaction we diagonalize the noninteracting Hamiltonian introducing new spin operators similar to Eq. (\ref{eq:new_spins}). 
The effective Hamiltonian reads 
\begin{widetext}
\begin{eqnarray}
\widehat{H}_{triple1}=\widehat{H}_{0}+\widehat{U},
\widehat{H}_{0}=-\sum_{i=1}^{3}E_{i}s_{i}^{z}, ~ \widehat{U}=-\frac{1}{2}\sum_{i,j=1}^{3}U_{ij}\frac{\Delta_{j}s_{i}^{z}-\Delta_{0i}s_{i}^{x}}{E_{i}}\frac{\Delta_{j}s_{j}^{z}-\Delta_{0j}s_{j}^{x}}{E_{j}}. 
\label{eq:Htriple1}
\end{eqnarray} 
\end{widetext}
Assume that we have resonant transition involving all three TLS between states $s_{i}^{z}=s_{j}^{z}=-s_{k}^{z}=1/2 \leftarrow\rightarrow s_{i}^{z}=s_{j}^{z}=-s_{k}^{z}=-1/2$ or equivalently $|++-> \leftarrow\rightarrow |--+>$. In resonance the transition energy $E_{i}+E_{j}-E_{k}$ must be small so in our estimate we set $E_{i}+E_{j}\approx E_{k}$. The transition amplitude between states $|++->$ and $|--+>$ is defined by the TLS interaction $\widehat{U}$. In the first order in $\widehat{U}$ one has $J_{ijk}=<++-|\widehat{U}|--+>=0$. The first term is equal zero because there is no three spin transition terms in the TLS Hamiltonian Eq. (\ref{eq:Htriple1}) because such terms must contain the product of three $s^{x}$ terms, while the maximum number of $s^{x}$ terms is two. In the third order the transition amplitude can be expressed as 
\begin{equation}
J_{ijk}=-\sum_{a}\frac{<++-|\widehat{U}|a><a|\widehat{U}|--+>}{E_{a}-E_{i}-E_{j}-E_{k}}, 
\label{eq:triple_trans1}
\end{equation}
where the sum is taken over all possible intermediate states of the Hamiltonian $\widehat{H}_{0}$ Eq. (\ref{eq:Htriple1}). Taking all possible intermediate states we found the following final expression
\begin{eqnarray}
J_{ijk}\approx -\frac{\Delta_{0i}\Delta_{0j}\Delta_{0k}}{8E_{k}E_{j}E_{k}}
\nonumber\\
\times\left[-\frac{U_{ij}U_{ik}\Delta_{i}}{E_{j}E_{k}}-\frac{U_{ij}U_{jk}\Delta_{j}}{E_{i}E_{k}}+\frac{U_{ik}U_{jk}\Delta_{k}}{E_{i}E_{j}}\right]. 
\label{eq:triple_trans_ans}
\end{eqnarray}
This result fully disagrees with the estimate Eq. (\ref{eq:bodea}) of Ref. \cite{subject}. When the distance between, say, TLS $k$ and TLSs $i$ and $j$ becomes large the amplitude Eq. (\ref{eq:triple_trans_ans}) decreases as $J_{ijk} \propto 1/R_{ik}^{3} \sim 1/R_{jk}^{3}$ in contrast with Eq. (\ref{eq:bodea}) where this amplitude is independent of the longest distance between TLS.

\begin{figure}[b]
\centering
\includegraphics[width=8cm]{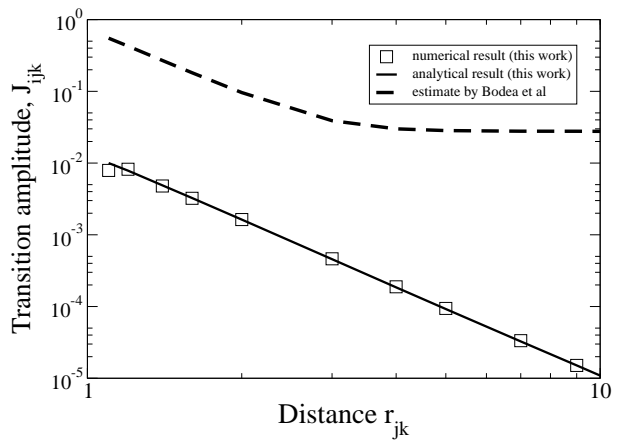}
\caption{Dependence of a triple tunnelling amplitude on the distance between TLS in different models.
  }\label{fig:1} 
\end{figure}

We also verified our result for the triple TLS transition amplitude using exact numerical solution of the Hamiltonian Eq. (\ref{eq:Htriple1}). 
The most straightforward definition of the transition amplitude is based on the repulsion of energy levels. If we have two nearly degenerate energy levels with energies $E_{1}$ and $E_{2}$ coupled by the perturbation matrix element $V$ then the energy splitting of two levels can be represented as 
\begin{equation}
\delta E = 2\sqrt{\frac{(E_{1}-E_{2})^{2}}{4}+V^{2}}. 
\label{eq:resonant_splitting}
\end{equation}
Then the coupling matrix element can be defined as the minimum energy splitting for two energy levels. One can extract the transition matrix element $V$ from the exact solution for the energy levels of the Hamiltonian Eq. (\ref{eq:Htriple1}). We minimize the splitting of energy levels with respect to the asymmetry energy of the third TLS $\Delta_k$. Generally one can find four local energy minima corresponding to two pair resonances $E_{k}\approx E_{i}$ and $E_{k}\approx E_{j}$ and two triple resonances, which are realized if $E_{i}+E_{j}\approx E_{k}$ and $|E_{i}-E_{j}|\approx E_{k}$. We are not interested in pair resonances because their number has been already estimated and proved to be small.\cite{BMP,subject} We study only one of the two pair resonances corresponding to the condition  $E_{i}+E_{j}\approx E_{k}$, which was assumed in our analytical estimate Eq. (\ref{eq:triple_trans_ans}) using the Scilab package.\cite{scilab}. 
We have generated the parameters of three TLSs using the random number generator between $0$ and $1$ which are $\Delta_{i}=0.544$, $\Delta_{j}=0.232$,
$\Delta_{0i}=0.231$, $\Delta_{0j}=0.216$ and $\Delta_{0k}=0.883$, take the displacement vector between TLSs $i$ and $k$ to be $(0, 0, 2)$ and compute the transition amplitude as the function of the length of the vector connecting TLS $i$ and $j$ in the form ${\bf r_{jk}}=\eta (2 0 0)$. The interaction has been set in the form $U_{ij}=1/r_{ij}^3$; $U_{ik}=-1/r_{ik}^3$; $U_{jk}=0.5/r_{jk}^3$. The graph in Fig. \ref{fig:1} shows the comparison of numerical approach with our predictions Eq. (\ref{eq:triple_trans_ans}) and the predictions of  Eq. (\ref{eq:bodea}), Ref. \cite{subject}. It is clear that the approach of Ref. \cite{subject} deviates from our estimate and exact calculations by several orders of magnitude. 
The analysis of the second resonance leads to the same results. Thus we minimize the energy splitting between two energy levels of the Hamiltonian Eq. (\ref{eq:Htriple1}) assuming 
$E_{i}+E_{j}\approx E_{k}$. If we enumerate all eight energy levels in ascending order these would be fourth and fifth energy levels having the energies $E_{4,5}\approx \pm(E_{i}+E_{j}-E_{k}) \approx 0$.  The minimization leads to the resonant asymmetry energy $\Delta_{k*}$. 
 
One can estimate the number of triple resonances $W_{3}$ using the transition amplitude Eq. (\ref{eq:triple_trans_ans}). This number can be obtained summing the number of all TLS triples involving the given TLS with the constraint that the energy of TLS transition $\mid E_{i}+E_{j}-E_{k}\mid$ is less than the transition amplitude $J_{ijk}$ Eq. (\ref{eq:triple_trans_ans}). This constraint can be approximately taking into account by replacing the integral over the third TLS energy with $\mid J_{ijk}\mid$. Then the number of triple resonances per the single TLS $i$ can be estimated as  
\begin{eqnarray}
W_{3}\approx P_{0}^{2}\int_{0}^{k_{B}T} d\delta_{j}\int\frac{d\Delta_{0j}}{\Delta_{0j}} \int\frac{d\Delta_{0k}}{\Delta_{0k}} \int d{\bf r}_{2}\int d{\bf r}_{3}J_{ijk}
\nonumber\\
\theta(\lambda_{T}-r_{1})\theta(\lambda_{T}-r_{2})\theta(r_{1}-a_{T})\theta(r_{2}-a_{T})
\nonumber\\
\approx P_{0}^{2}U_{0}^{2}\ln^{2}\left(\frac{k_{B}T}{\hbar \gamma_{ph}}\right)\approx W_{2}^2\ll 1,  
\label{eq:triple_res_est}
\end{eqnarray}
This result is different both qualitatively and quantitatively from the result of Ref. \cite{subject}.  Remind that the contribution of such TLS results in  the huge phase volume dimensionless factor $\frac{k_{B}T}{\hbar \gamma_{ph}}$ which results in the large number of triple resonances per one TLS in \cite{subject}. 

Thus according to our estimate, triples are not contribute significantly to the delocalization of TLS energy. This was already discussed in our previous work \cite{BMP}. If however the more complicated clusters of four two level systems will be taken into consideration the number of resonances will increase as the power of the effective size $R_{max} \sim $. This four particle clusters made of resonant pairs of two-level systems are responsible for the energy delocalization as described in the previous work.\cite{BMP} 

This work is supported by the Louisiana Board of Regents, (Contract No. LEQSF (2005-08)-RD-A-29) and by Tulane University Research Enhancement Awards (Phase 1 and Phase 2). We also acknowledge Philip Stamp and Clare Yu for stimulating discussions, useful comments and advices.

\end{document}